%% file: qcd_lep2.tex
\documentclass[12pt,a4paper,dvips]{article}
\usepackage{a4p}
\usepackage{cite,mcite}
\usepackage{graphicx}
\usepackage{rotating}
\usepackage{physics}
\usepackage{l3_title}
\usepackage{ifthen,Lep}
\journalname{Phys. Lett. B}
\date{February 13, 2002}
\preprint{2002-015}
\Lep{2}
\graphicspath{{.}}
\newlength{\capindent}
\newlength{\capwidth}
\newlength{\figwidth}
\newcommand{\icaption}[2][!*!,!]{\hspace*{\capindent}%
  \begin{minipage}{\capwidth}
    \ifthenelse{\equal{#1}{!*!,!}}%
      {\caption{#2}}%
      {\caption[#1]{#2}}
  \end{minipage}}
\def\as{\alpha_{\mathrm{s}}}

\def\bt{B_{T}}
\def\bw{B_{W}}
\def\ee{\mathrm{e^{+}e^{-}}}

\def\epar{E_{\parallel}}
\def\eperp{E_{\perp}}
\def\etal{{\it et~al.}}%
\def\evis{E_{{vis}}}
\def\WW{\mathrm{W^{+}W^{-}}}

\def\rs{\sqrt{s}}
\def\rsp{\sqrt{s'}}

\def\qq{\mathrm{q\bar{q}}}

\def\ncl{N_{{cl}}}

\def\pb{\mathrm{pb^{-1}}}

\begin{document}
\bibliographystyle{l3style}
\begin{titlepage}
\title{Determination of {\boldmath $\as$} from Hadronic Event Shapes
in {\boldmath $\ee$} Annihilation at 192 {\boldmath $\leq$}
                  {\boldmath $\rs$} {\boldmath $\leq$} 208 
      {\boldmath \GeV}}

\author{The L3 Collaboration}
%
%

\begin{abstract}

Results are presented from a study of the structure of 
high energy hadronic events recorded by the L3 detector at $\rs \ge$ 192 \GeV. 
The distributions of several event shape variables are compared
 to   resummed $\cal{O}$$(\as^{2})$  QCD calculations.
We determine the strong coupling constant at three average centre-of-mass 
energies: 194.4, 200.2 and 206.2 \GeV.
These measurements, combined with previous L3 measurements at lower energies,
demonstrate the running of $\as$ as expected in QCD 
and yield $\as (\MZ)$ = 0.1227 $\pm$ 0.0012  $\pm$ 0.0058, where the first uncertainty is
experimental and the second is theoretical.
\end{abstract}
\submitted
\end{titlepage}

\normalsize

\section{Introduction}\label{sec:intr}

The measurement of the energy dependence of the strong coupling constant, $\as$, is an important 
test of Quantum Chromo-Dynamics (QCD).
Hadronic events produced in  $\ee $ annihilation offer a  clean 
environment to perform such measurements. 
The high energy phase of the LEP collider gives a unique opportunity to 
measure QCD observables over a wide energy range and to perform a precise
test of the energy dependence of the strong coupling constant.
In addition, the study of hadronic events
 allows to check the validity of the QCD models 
used for background modelling
in other studies, such as new particle searches.

In its last two years,  the LEP collider operated at various centre-of-mass energies, $\rs$,
between 192 and 208 \GeV. These are grouped in three samples of average 
centre-of-mass energies  194.4, 200.2 and 206.2 \GeV, corresponding to the
ranges 192 $\le\rs<$ 196 \GeV, 200~$<\rs<$~202~\GeV\ and $\rs>$ 202 \GeV\ respectively.

We report on measurements of five event shape distributions 
using 436.8 $\pb$ of data collected with the L3 detector~\cite{l3:det} at these 
centre-of-mass energies, as detailed in Table~\ref{tab:events}. To allow a direct comparison with our earlier QCD studies 
at lower energies
\cite{l3qcd91,l3qqg,l3qcd133,l3qcd161-183,l3qcd189}, we follow an identical analysis 
procedure. 
The value of $\as$ is extracted in each energy range 
by comparing the measured event shape 
distributions with predictions of second order QCD calculations 
\cite{qcdz} supplemented by resummed leading and next-to-leading order terms 
\cite{qcdresum,oldbt,logR,newbt}. 
These values are used, together with  previous L3 measurements at
 lower effective centre-of-mass energies, from 30 \GeV\ to 189 \GeV, to study
the energy evolution of $\as$.

\section{Event Selection}\label{sec:sel}

The criteria for the selection of $\ee$ $\rightarrow$ $\qq$ $\rightarrow hadrons $ events  
are identical to those used in our previous QCD 
study at $\rs$ = 189 \GeV\ \cite{l3qcd189}. They are based on
the  measured total visible energy,  $\evis$ ,
the energy imbalances parallel, $\epar$, and perpendicular, 
 $\eperp$, to the beam direction and on the cluster multiplicity, $\ncl$. 
These variables are constructed using energy clusters in  
the electromagnetic and hadronic calorimeters with a  minimum energy of 100 \MeV.

The efficiency of the selection 
criteria and the purity of the data sample are estimated using Monte Carlo events for the process 
$\ee \rightarrow \qq (\gamma )$ generated by the KK2F 
\cite{kk2f} program, interfaced with JETSET PS~\cite{jetset}
routines to describe the QCD parton shower evolution and hadronisation. The events are then passed 
through the L3 detector simulation~\cite{l3-simul}. The  KK2F 
generator is chosen for its improved simulation of the initial state
radiation (ISR) as compared to the  PYTHIA~\cite{jetset} model previously used.  
Background events are simulated with PYTHIA 
  for two-photon events and   Z-pair production, 
 KORALZ~\cite{koralz} for the $\tau^{+}\tau^{-}(\gamma)$ final state, 
 BHAGENE~\cite{bhagene} and BHWIDE~\cite{bhwide} for Bhabha 
events and KORALW~\cite{koralw} for W-pair production.

Hadronic events with hard ISR photons, where the mass of 
the  hadronic system is close to $\MZ$, are considered as background if the photon energy 
exceeds 0.18$\rs$.
This important background is reduced to less than  8\% of the selected events 
by applying a cut in 
the  plane of $\mid$$\epar$$\mid$/$\evis$ $vs.$ $\evis /\rs$. 
Additional background arises from W-  and Z-pair production. 
A substantial fraction of these events is removed by a specific selection~\cite{l3qcd161-183}
that forces  
 a 4-jet topology using the Durham algorithm~\cite{kt}, and applies 
cuts on the cluster multiplicity, $\ncl>$ 40, the jet resolution parameter, $y_{34}^{D}>$0.0025, 
 the energy of the  least energetic jet and on the energy fraction carried by the
 two most energetic jets.
The  cuts are optimised to maximise the product
of efficiency and purity at each energy point.
After selection, the W-pair background amounts to  6.4\% at $\rs = 192 \GeV$ and increases
 to 10.3\% at $\rs =208 \GeV$.
The Z-pair background is below 0.8\%.
The   selection efficiency, purity and number of selected
events for the three  energy points are summarised in Table~\ref{tab:events}.

\section{Measurement of Event Shape Variables}\label{sec:evshap}

The measured global event shape variables 
 are the thrust~\cite{thrust}, $T$, the scaled heavy jet mass~\cite{hjm}, $\rho$, the total, $\bt$, and 
wide, $\bw$, jet broadening variables~\cite{oldbt} and the 
$C$-parameter~\cite{tmatx}. The first four observables are defined in terms of the particle
four-momenta, while  the $C$-parameter is derived from the spherocity tensor:
$${
\theta^{ij} = {{\sum_a {p^i_a p^j_a}/\mid\vec{p_a}\mid}\over 
               {\sum_a\mid\vec{p_a}\mid}}\ \  i,j=1,2,3\ ,} $$
where the sums run over all particles and 
${\vec{p}_{a}}$ is the momentum  vector of the particle $a$.
The $C$-parameter is defined in terms of the eigenvalues,  $\lambda_1$, 
$\lambda_2$ and $\lambda_3$, of $\theta^{ij}$, as:
$${
C = 3(\lambda_1\lambda_2 + \lambda_2\lambda_3 + \lambda_3\lambda_1)\ .}$$

For all   five variables, improved analytical QCD 
calculations \cite{qcdresum,oldbt,logR,newbt} are available.
The calculations used here for the jet broadening variables~\cite{newbt}
are improved as compared to the previous predictions~\cite{oldbt} by a better
treatment of quark recoil effects.

After  background subtraction, the measured 
distributions are corrected bin-by-bin for detector effects, acceptance and resolution.
 The correction factors are the ratios of Monte Carlo distributions
 at  detector level 
  to the distributions at particle level which include all stable
charged and neutral particles.~\footnote{All weakly decaying light 
particles with mean lifetime larger than $3.3\times 10^{-10}$ s are considered
  stable.}  
The data are also corrected bin-by-bin for initial and final state photon radiation
using Monte Carlo distributions at particle level with and 
without radiation. 

\begin{figure}[htbp]
\begin{center}
 \includegraphics[width=0.48\textwidth]{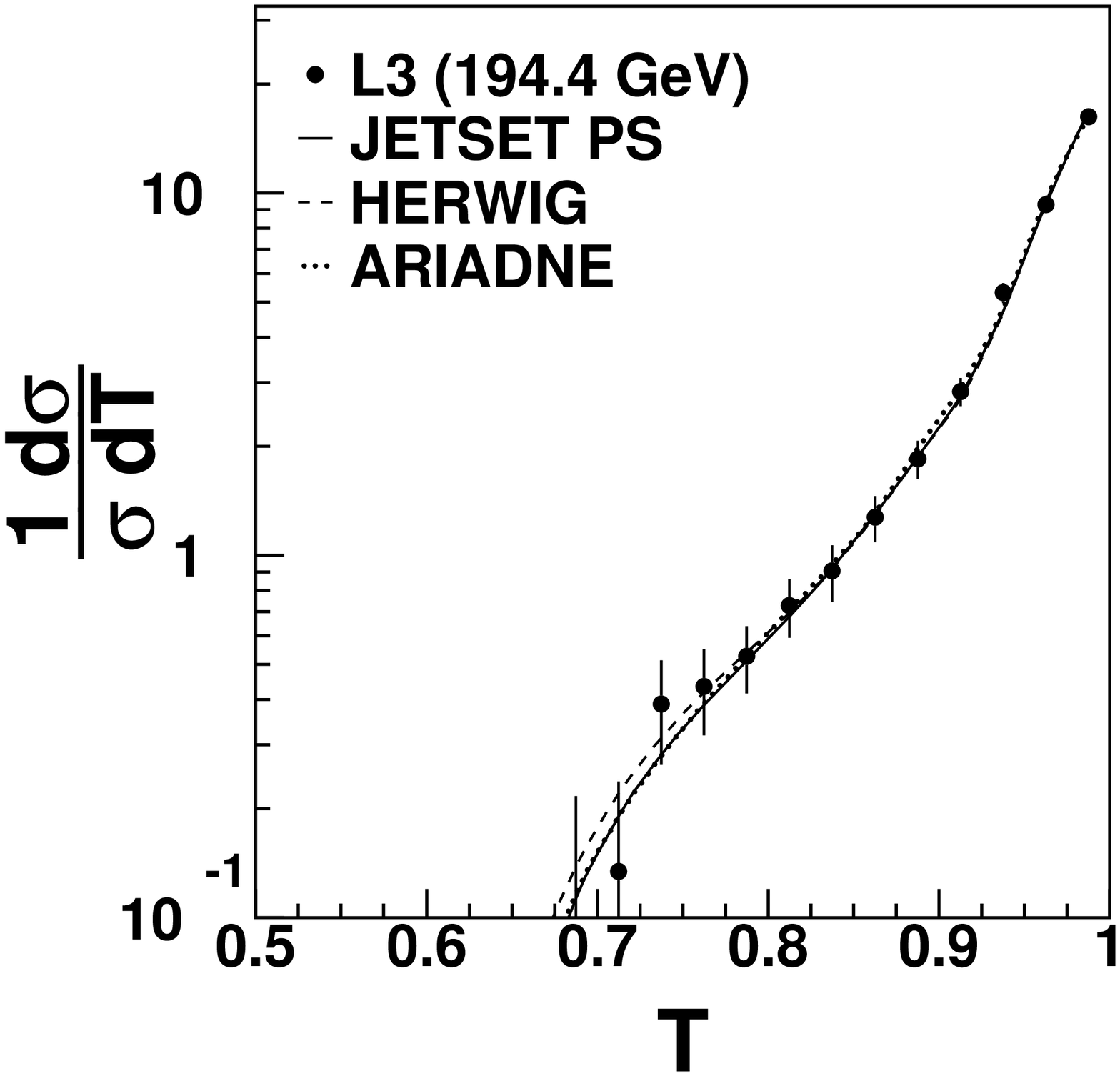}
 \includegraphics[width=0.48\textwidth]{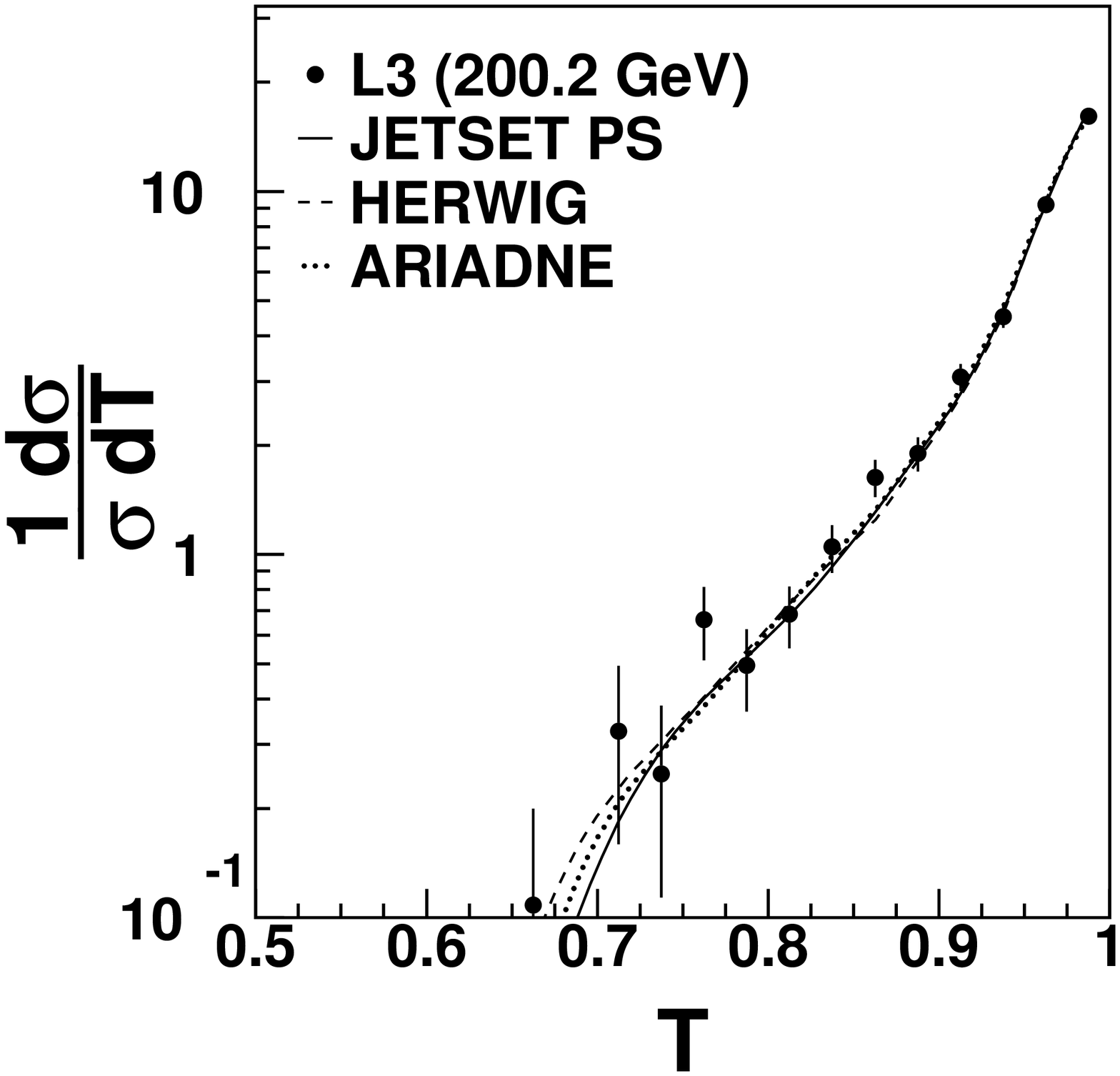}\\
 \includegraphics[width=0.48\textwidth]{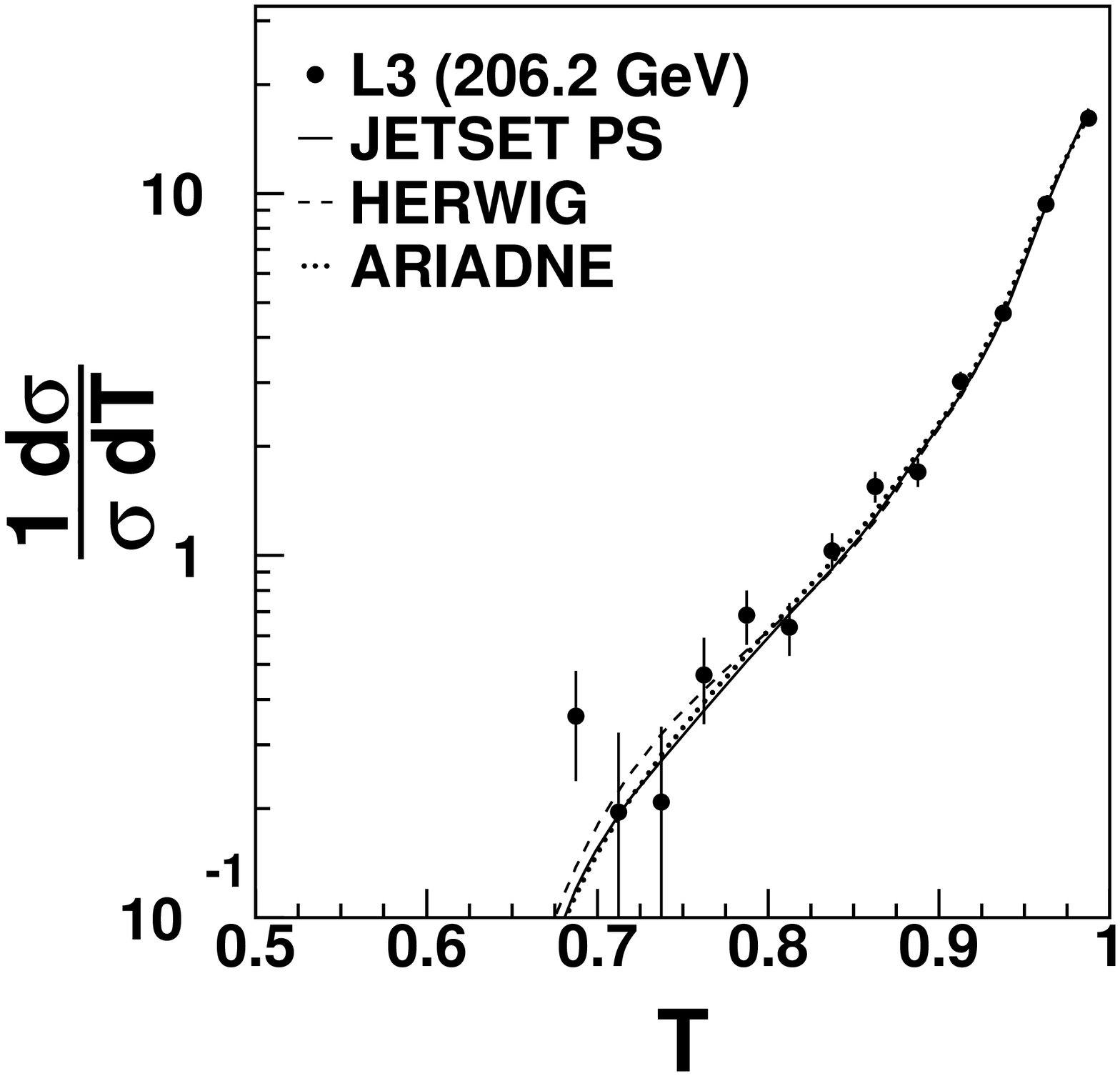}
\end{center}
\caption[]{Corrected distributions for the thrust, $T$, at $<\rs>$= 194.4, 200.2 and 206.2 
           \GeV\, compared with QCD model predictions. The uncertainties shown 
           are statistical only.}
\label{fig:thrust}
\end{figure}

\begin{figure}[htbp]
\begin{center}
 \includegraphics[width=0.48\textwidth]{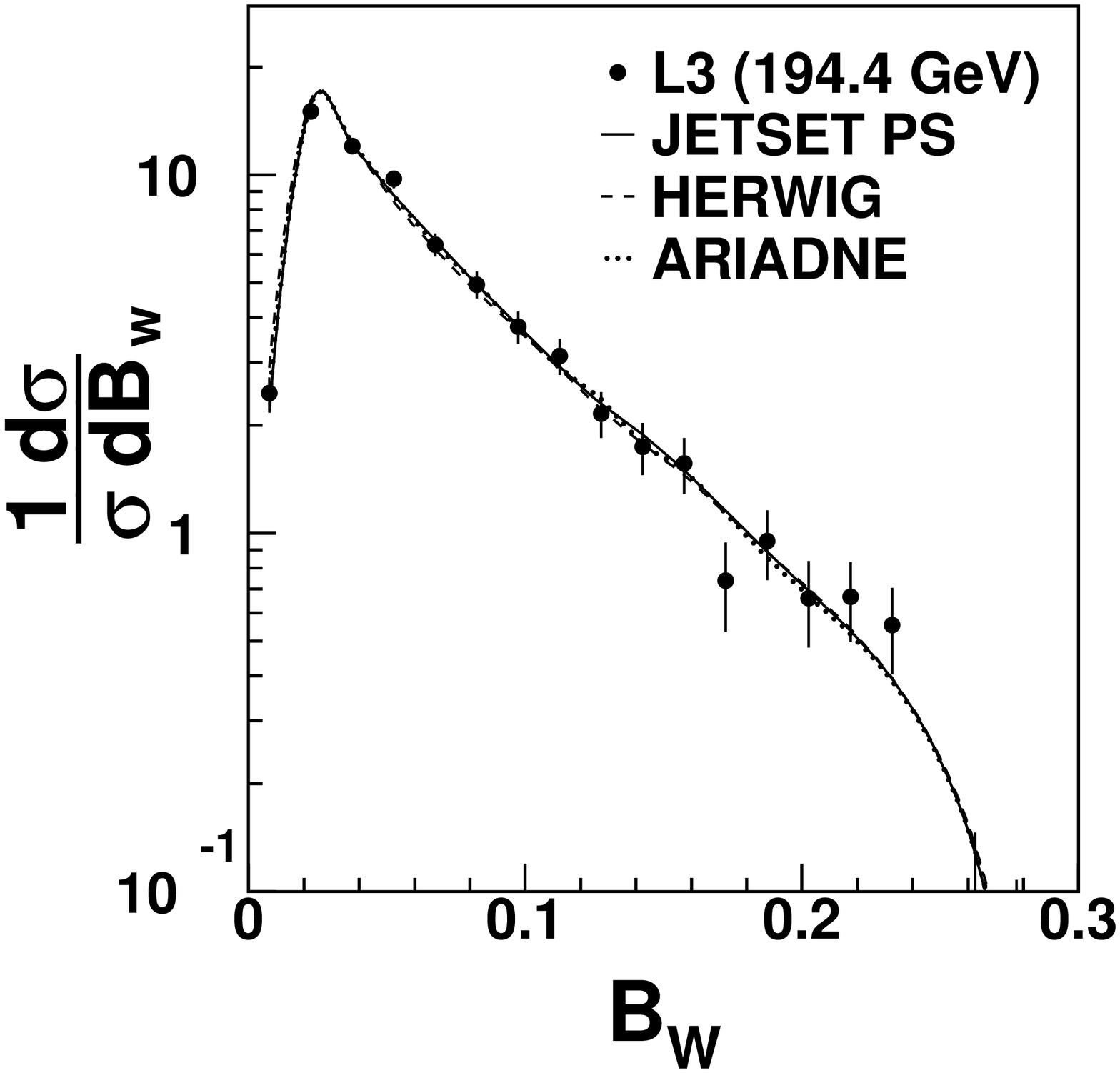}
 \includegraphics[width=0.48\textwidth]{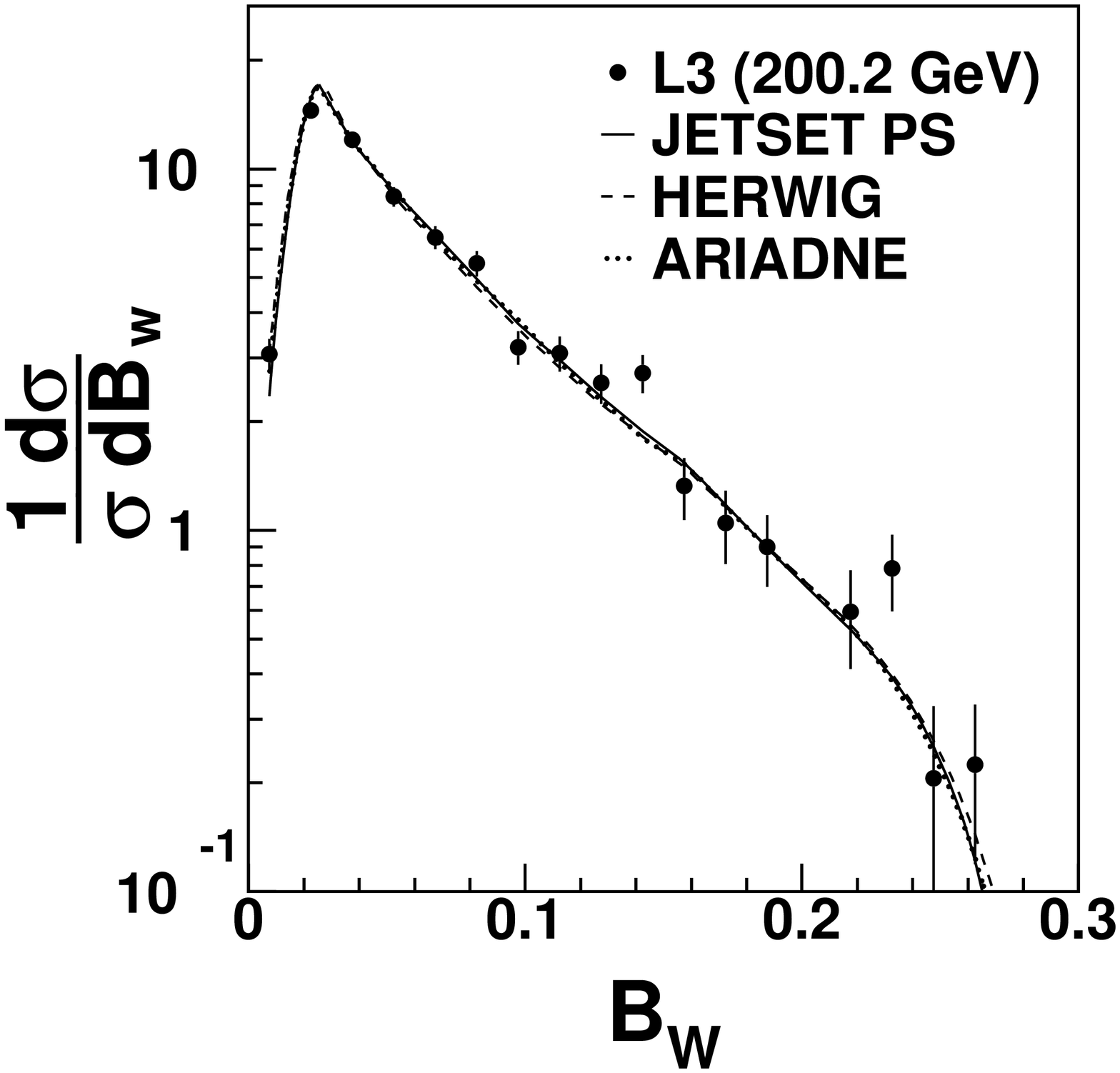}
 \includegraphics[width=0.48\textwidth]{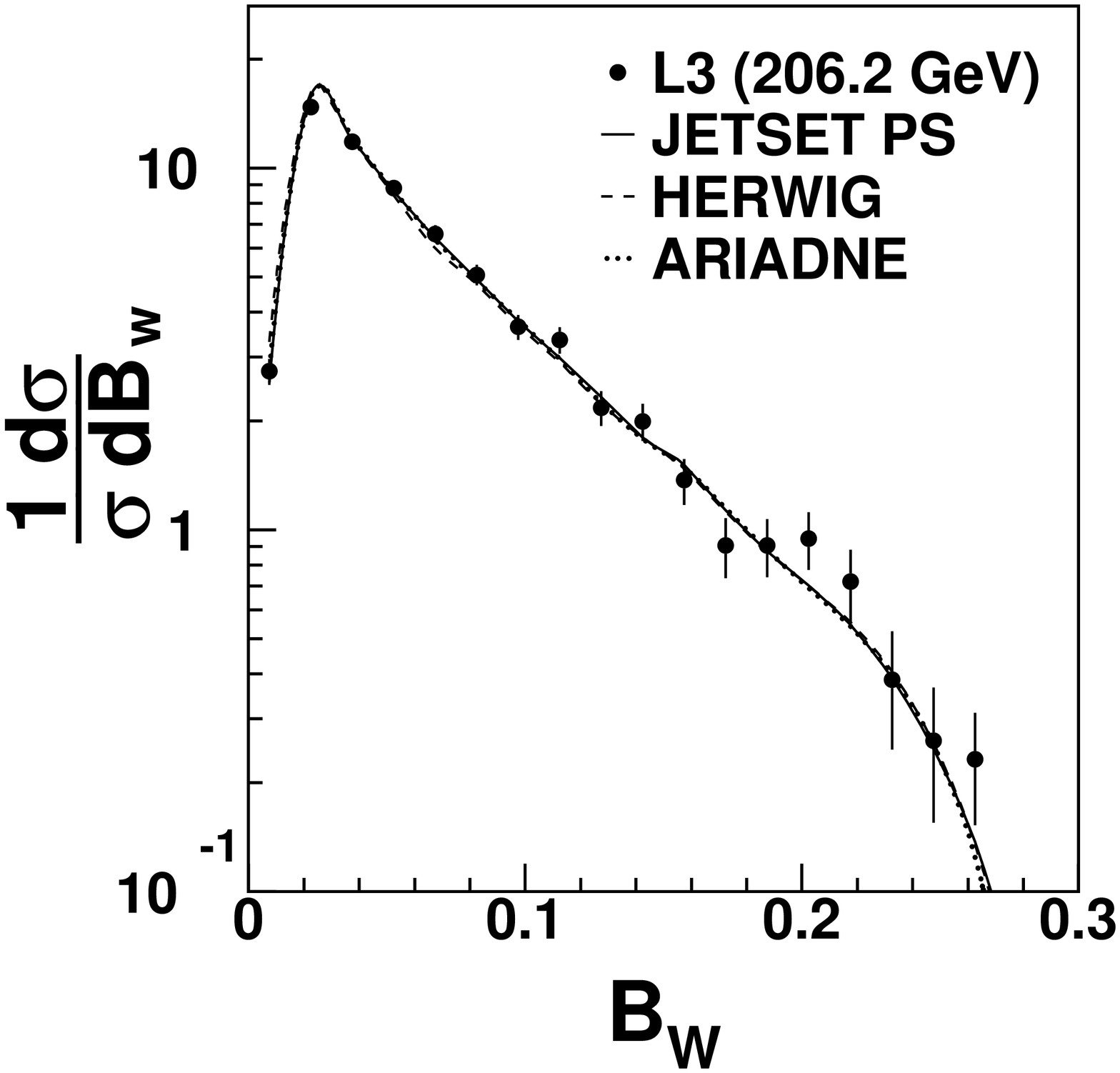}
\end{center}
\caption[]{Corrected distributions for the wide jet broadening, $\bw$,
           at $<\rs>$=  194.4, 200.2 and 206.2 \GeV\, compared with QCD 
           model predictions. The uncertainties shown are statistical only.}
\label{fig:bw}
\end{figure}

Figures \ref{fig:thrust} and \ref{fig:bw} show the  thrust and
wide jet broadening distributions corrected to the particle level.
The data are compared with the JETSET,  HERWIG~\cite{herwig} and 
 ARIADNE~\cite{ariadne} QCD models. These models, 
based on an improved leading-logarithmic approximation parton 
shower,  including QCD coherence effects, 
are tuned to reproduce the global event shape distributions
and the charged particle multiplicity distribution 
measured at 91.2 \GeV\ \cite{l3tune}. 
At and above centre-of-mass energies of 200.2 \GeV\, 
 some 
discrepancies appear for specific values of the observables.
Several studies are
 performed to investigate the reason for such effects.
The observed structures in the global event shape distributions are found to depend neither on time nor on detector geomety.
The effects of these discrepancies  are taken into account in the determination of the systematic
uncertainties.

The two main sources of systematic uncertainty  in the event shape variable distributions 
are those on  detector correction and background estimation.
These uncertainties are estimated by repeating the measurement with different  
analysis criteria and correction procedures~\cite{l3qcd161-183,l3qcd189}.
The uncertainty in the detector correction is estimated with the following tests:
\begin{itemize}
\item The effect of different particle fluxes in correcting the measured
       distribution is estimated by using the HERWIG   Monte Carlo
       program instead of JETSET to simulate the signal.
       Half of the difference obtained with these two models is taken as a systematic uncertainty.
      
 \item The definition of reconstructed objects used to calculate the 
       observables is changed from calorimetric  clusters to a 
       non-linear combination of energies of charged tracks and calorimetric 
       clusters.

 \item The acceptance is reduced by restricting the analysis to events in the 
       central part of the detector, $|\cos(\theta_{T})|< 0.7$, where 
       $\theta_{T}$ is the polar angle of the thrust axis relative 
       to the beam direction. In this region a better energy resolution is found.
\end{itemize}
       Half of the  maximum spread between the latter two tests
       and the original analysis
        is assigned as a  systematic uncertainty.

The uncertainty on the background composition of the selected  sample
is estimated by repeating the analysis with:
\begin{itemize}
 \item an alternative criterion to reject the hard initial state photon 
       events based on a cut on the kinematically reconstructed effective 
       centre-of-mass energy ($\rsp<0.92$);
\item a variation of the $\WW$ background estimate by applying a full subtraction of
 the  W-pair contribution without preliminary event rejection;
 \item a variation of the estimated two-photon interaction background by 
       $\pm$ 30\%. The  Monte Carlo program used to model two-photon interactions is 
       also changed
       from PYTHIA to PHOJET~\cite{phojet}.
\end{itemize}
 For the first two studies, half of the difference between the results of the
 original analysis and of the systematic check is taken as the systematic uncertainty. 
 In the two-photon case, half of the maximum spread between the new results and the original 
 analysis is considered as a systematic uncertainty.
The statistical component of each systematic uncertainty is 
 estimated and removed following the  procedure described in reference~\cite{l3qcd189}.
The  systematic uncertainties  obtained from  the different sources are 
then combined in quadrature.
For $\rs < 196 \GeV$, the uncertainties due to the backgrounds are the most 
important ones. They are $2-3$ times
larger than the uncertainties due to detector corrections.
For  $\rs > 200 \GeV$,  the uncertainty in the 
detector correction gives the largest systematic contribution, dominated by the
effect of reducing the event thrust acceptance in the central part of the detector, but decreases
for $<\rs> = 206.2 \GeV$. 

\begin{figure}[htbp]
\begin{center}
 \includegraphics[width=0.7\textwidth]{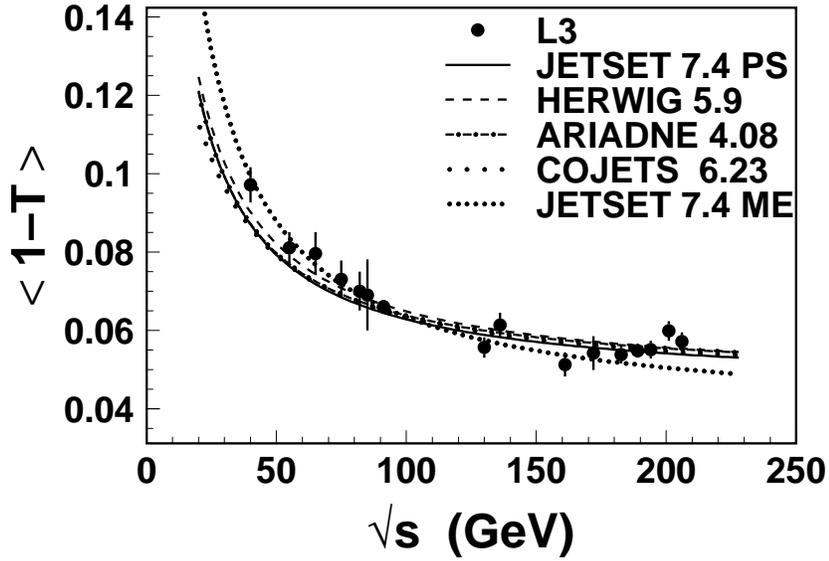}
 \includegraphics[width=0.7\textwidth]{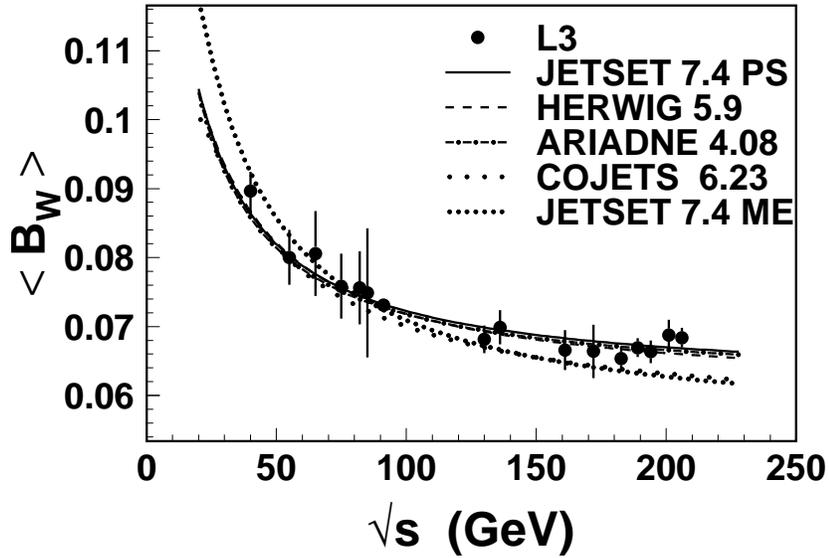}
\end{center}
\caption[]{Distribution of a) $<1-T>$ and b)
 $<\bw>$ as a function of the centre-of-mass energy, compared to several
 QCD models. Lower energy data~\cite{l3qcd91,l3qqg,l3qcd133,l3qcd161-183,l3qcd189}
  are also presented.
   The error bars  include  experimental systematic uncertainties.}
\label{fig:evol}
\end{figure}

An important test of QCD models
is the comparison of the energy
evolution of the means of the event shape variables.
The mean values of the five variables obtained at $\rs >$ 192 \GeV\
are given in Table~\ref{tab:fmom}.
 Figure~\ref{fig:evol} shows the evolution 
of $<1-T>$ and $<\bw>$ as a function of $\rs$.
Also shown are the energy dependences of these quantities as predicted by
JETSET PS, HERWIG, ARIADNE, COJETS~\cite{cojets} and JETSET ME, with an $\cal{O}$($\as^{2}$) 
matrix element implementation.
For high energies, the JETSET ME and COJETS models are not favoured by the data.

\section{Determination of {\boldmath $\as$}} 

The QCD predictions for the five event shape observables are
 based on $\cal{O}$$(\as^{2})$ perturbative
QCD calculations with resummed leading and next-to-leading order terms.
To compare these   calculations at parton level 
with the experimental distributions,
the effects of hadronisation and decays are corrected for with a folding matrix~\cite{l3qcd133}
calculated using the JETSET PS Monte Carlo program.

\begin{figure}[htbp]
\begin{center}
    \includegraphics*[width=16.0cm]{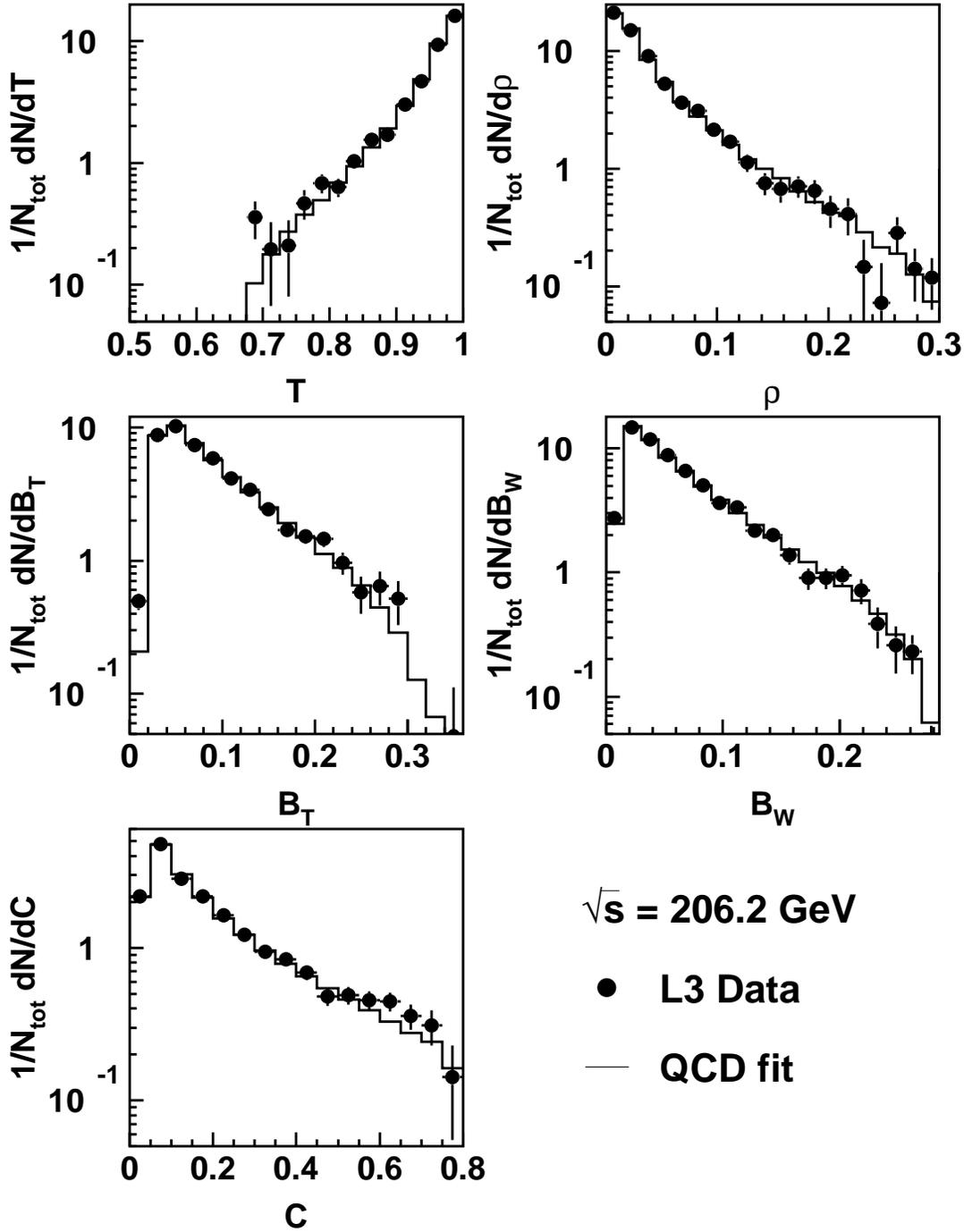}
\end{center}
\caption[]{Measured distributions of thrust, $T$, scaled heavy jet mass, 
           $\rho$, total, $\bt$, and wide, $\bw$, jet broadenings, and
            $C$-parameter in comparison with QCD predictions at $<\rs>$=206.2 $\GeV$. 
              The error bars  include  experimental systematic uncertainties.}
\label{fig:alsfit}
\end{figure}

To determine $\as$      at each energy point,
the measured distributions  are fitted in the ranges  given in Table~\ref{tab:als} 
to the analytical predictions,
using the modified-log(R) matching scheme~\cite{logR} after corrections for hadronisation effects.
Figure~\ref{fig:alsfit} shows the experimental data together
with the result of the QCD fits for the five variables 
at $<\rs>$ = 206.2 \GeV. 

The $\as$ measurements at the three energy points 
are summarised in Table~\ref{tab:als} together with their experimental and theoretical uncertainties.
The former includes the statistical and
the experimental systematic uncertainties discussed above.
The latter is obtained from estimates~\cite{l3qcd133} of the 
hadronisation uncertainty and of  the uncalculated 
higher orders in the QCD predictions.

The hadronisation uncertainty is obtained from the
 variation in the fitted value of $\as$ due to
 hadronisation corrections determined 
by comparing JETSET with HERWIG and ARIADNE models
and changing the JETSET fragmentation parameters, $b$, $\sigma_{q}$ and
$\Lambda_{LLA}$ within their errors~\cite{l3tune} and turning off Bose-Einstein correlations. 
The most important variation comes from the change in the fragmentation models and is taken as 
an estimate of the overall hadronisation uncertainty.

The uncertainty  coming
from uncalculated higher orders in the QCD predictions is estimated in two independent ways: by
varying the renormalisation scale, $\mu$, and by changing the matching scheme.
The scale uncertainty is obtained by repeating the fit for different values of the renormalisation scale
in the interval $0.5 \rs \leq \mu \leq  2 \rs$.
The matching scheme uncertainty is obtained from half of the
maximum spread given by   different algorithms~\cite{logR}.
The largest of these uncertainties  is assigned as the theoretical
uncertainty due to uncalculated higher orders.

To obtain a combined value for the strong coupling constant, we take the 
unweighted average of the five $\as$ values.
The overall theoretical uncertainty is obtained from the 
average  hadronisation uncertainty added in quadrature to the average 
higher order uncertainty.
 A cross-check
of this theoretical uncertainty is obtained from a comparison of $\as$ 
measurements from the various event shape variables which are
expected to be differently affected by higher order corrections and hadronisation effects.
Half of the maximum spread in the five $\as$ values is found to be
consistent with the estimated theoretical uncertainty.

Earlier L3 measurements at $\rs$ = $\MZ$ and at reduced centre-of-mass 
energies determined $\as$ from four event shape variables only: $T$, $\rho$, 
$\bt$ and $\bw$, the resummed calculation for the $C$-parameter not being  
available. We have determined $\as$ at these
lower energies from the $C$-parameter and the values are now 
included in the overall mean $\as$ and listed in Table~\ref{tab:alscomb1}.

The improved  theoretical predictions for the jet broadening variables
 are  used to update our previously 
 published $\as$ results at effective centre-of-mass energies from 30 \GeV\ up to 189 \GeV,
 as listed in Table~\ref{tab:alscomb1}. 
The  mean $\as$ values from  the five 
event shape distributions are given in Table~\ref{tab:alscomb2}
together with the experimental and theoretical uncertainties.

\begin{figure}[htbp]
\vspace*{-5cm}
\begin{center}
    \includegraphics[width=14cm]{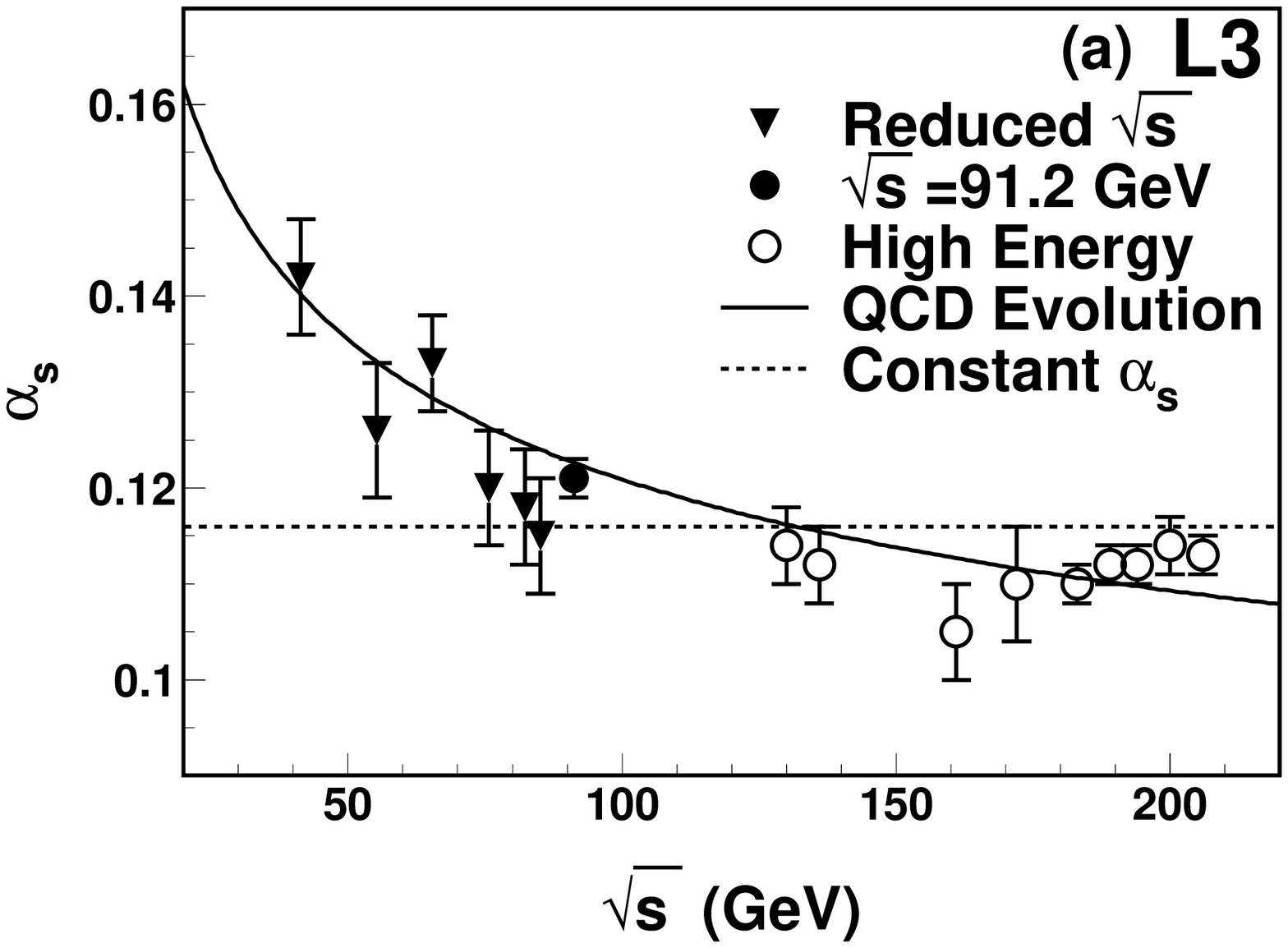}
    \includegraphics[width=14cm]{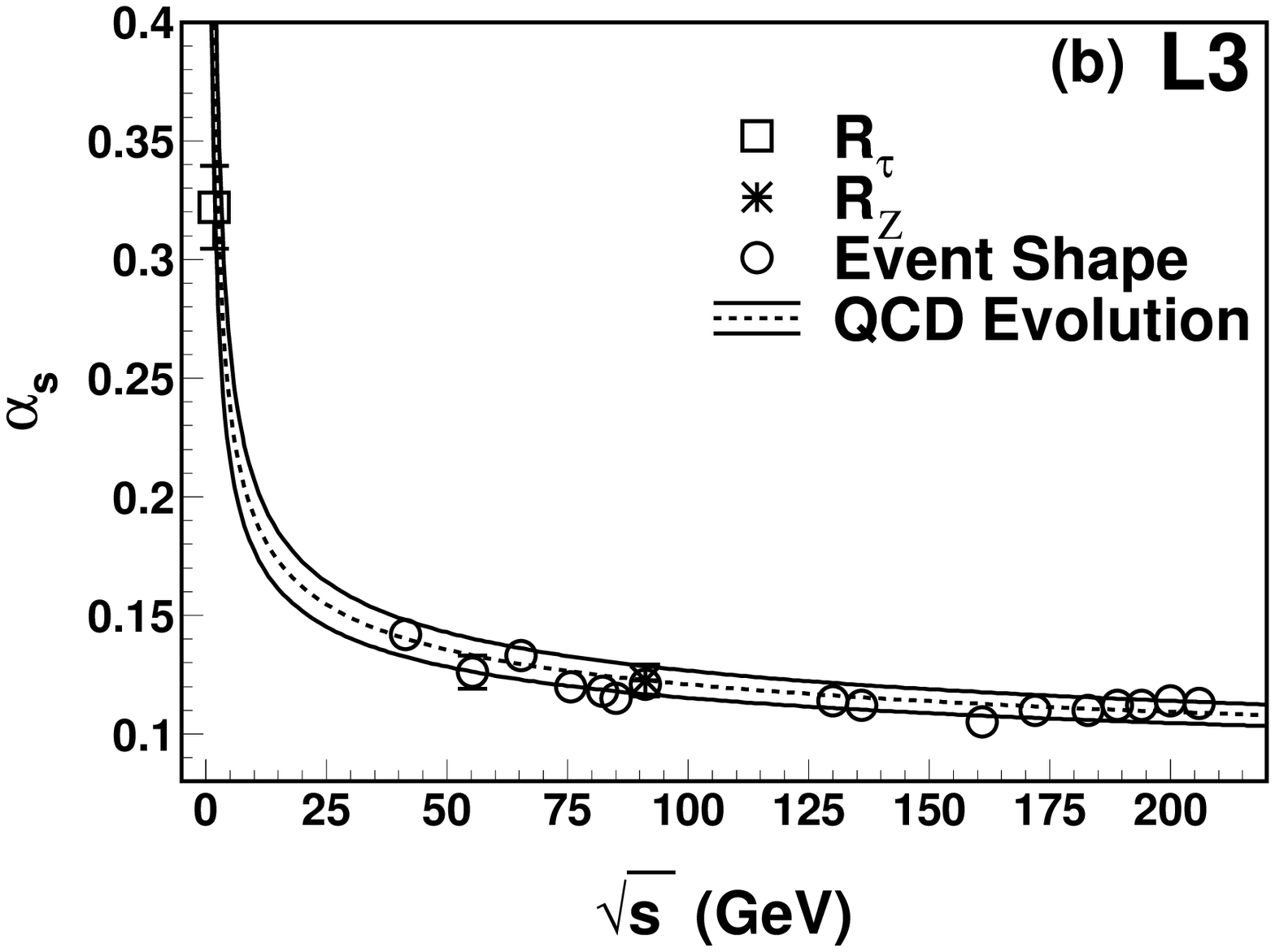}
\end{center}
\caption[]{Values of $\as$ determined as a function of $\rs$:  
           a) from event shape distributions with  experimental uncertainties
          only. The solid and dashed lines are fits with the energy dependence
          of $\as$ as expected from QCD and with constant $\as$, respectively;
          b) from the measurement of the $\tau$ branching fractions into
          leptons~\cite{l3tau},
           Z lineshape~\cite{l3lineshape} and event shape distributions. The dashed line is a
          fit to the QCD evolution function to the measurements made from
          event shape variables. The band width  corresponds to the evolved uncertainty on $\as(\MZ)$.
         }
\label{fig:alsevol}
\end{figure}

Figure \ref{fig:alsevol}a compares the energy dependence of 
the measured $\as$ values
with the prediction from QCD.
 The theoretical uncertainties are strongly correlated between these measurements. 
Hence, the energy dependence of 
$\as$ is investigated using only experimental uncertainties. 
The experimental systematic uncertainties on $\as$ are partially correlated. 
The background
uncertainties are similar for  data points in the same energy range
  but differ between the low energy, Z peak and high energy
data sets. 
The sixteen measurements in Figure~\ref{fig:alsevol}a are shown with 
experimental uncertainties only, together with a fit to the QCD evolution equation
\cite{pdg} with $\as(\MZ)$ as a free parameter, that takes
 into account the correlation between the various measurements.  
The covariance matrix for the fit is  obtained as follows: 
\begin{itemize}
\item The statistical uncertainties are taken as uncorrelated.
\item    The experimental systematic uncertainties are assumed to be 
  uncorrelated between the three data sets 
 and to have a minimum overlap correlation 
  between different energies within the same data set. This definition consists of
  assigning to the covariance
  matrix element  the smallest of the two squared uncertainties. 
\end{itemize}

The fit gives a $\chi^2$ of 17.9 for 15 degrees of
freedom corresponding to a confidence level of 0.27 yielding a value of $\as$:
\begin{eqnarray*}
  \as (\MZ) & = & 0.1227~\pm~0.0012~\pm~0.0058 \; .
\end{eqnarray*}

The first uncertainty is experimental and the second   theoretical. The latter
is obtained from the result of a fit which includes the theoretical uncertainties 
and their correlations. The covariance matrix is here defined  assuming a 
 minimum overlap correlation  between energies of the hadronisation as well as the uncalculated 
 higher order uncertainties. The  
  hadronisation uncertainty contribution to the total theoretical uncertainty 
  is $\pm0.0026$. 

A fit with constant $\as$ gives a $\chi^2$ of 51.7
for 15 degrees of freedom.
These measurements support the energy evolution of the strong
coupling constant predicted by QCD. 
The apparent increase of 
 the $\as$ values obtained  at $\rs >$ 194 \GeV\ compared to the QCD evolution curve is related
  to the structures  seen in the event shape distributions  discussed above. 
   
Figure~\ref{fig:alsevol}b summarises the $\as$ values determined by L3  
from the measurement of the $\tau$ branching fractions into leptons~\cite{l3tau}, 
Z lineshape~\cite{l3lineshape}
and event shape distributions at various energies, together with the QCD 
prediction obtained from the fit to the event shape measurements only. The band width 
corresponds to the evolved uncertainty on $\as(\MZ)$.
All the measurements are consistent with the energy evolution of the strong
coupling constant predicted by QCD.
The uncertainties on these measurements are dominated by the theoretical uncertainty coming
from the unknown higher order contributions in the calculations. 
An improved determination of $\as$ from these measurements thus awaits improved theoretical
calculations of these corrections.
 
%
%
\newpage
\section*{Author List}
\input namelist251.tex
\newpage

\newpage

\begin{table}[htbp]
\begin{center}\begin{tabular}{|l|c|c|c|c|}\hline
$<\rs>$ (GeV)                 & 194.4  & 200.2  & 206.2  \\ \hline
Integrated Luminosity ($\pb$) & 112.2  & 117.0  & 207.6 \\ \hline
Selection Efficiency (\%)     &  82.8  & 85.7   & 86.0 \\ \hline
Sample Purity (\%)            &  81.4  & 80.6   & 78.8 \\ \hline
Selected Events               &  2403  & 2456   & 4146 \\ \hline
\end{tabular}\end{center}
\caption[]{Summary of integrated luminosity, selection efficiency, sample 
           purity and number of selected  hadronic events at the 
           average centre-of-mass energies used in this analysis.}
\label{tab:events}
\end{table}

\begin{table}[htbp]
\begin{center}
\begin{tabular}{|c|c|c|c|}\hline
 $<\rs>$  &  194.4 \GeV & 200.2 \GeV & 206.2 \GeV \\ \hline
$<(1-T)>$ & 
  0.0551  $\pm$  0.0021  $\pm$  0.0009 &  
  0.0582  $\pm$  0.0021  $\pm$  0.0015 & 
  0.0569  $\pm$  0.0017  $\pm$  0.0016 \\ \hline
$<\rho>$&
  0.0439  $\pm$  0.0014  $\pm$  0.0007 &  
  0.0464  $\pm$  0.0014  $\pm$  0.0015 & 
  0.0455  $\pm$  0.0011  $\pm$  0.0011 \\ \hline
$<\bt>$ &
  0.0920  $\pm$  0.0022  $\pm$  0.0023 &  
  0.0950  $\pm$  0.0021  $\pm$  0.0025 & 
  0.0938  $\pm$  0.0017  $\pm$  0.0015 \\ \hline 
$<\bw>$ &
  0.0663  $\pm$  0.0014  $\pm$  0.0009 &  
  0.0688  $\pm$  0.0013  $\pm$  0.0016 & 
  0.0682  $\pm$  0.0011  $\pm$  0.0009 \\  \hline
$<C>$   &
  0.2158  $\pm$  0.0058  $\pm$  0.0023 &  
  0.2244  $\pm$  0.0059  $\pm$  0.0068 & 
  0.2195  $\pm$  0.0049  $\pm$  0.0035 \\  \hline
\end{tabular}
\end{center}
\caption[]{Mean values of the five event shape variables at different
           energy points. The first  uncertainty is statistical 
           and the second systematic.}
\label{tab:fmom}
\end{table}

\begin{table}[htbp]
\begin{center}
{\small
\begin{tabular}{|l|c|c|c|c|c|}\hline
          & $1-T$ & $\rho$ & $\bt$ & $\bw$ & $C$ \\ \hline
Fit Range         & 0.00$-$0.25 & 0.00$-$0.20 & 0.02$-$0.26 
                  & 0.015$-$0.210 & 0.05$-$0.50 \\ \hline\hline
$\as$(194.4 \GeV)    & 0.1168 & 0.1096 & 0.1152 & 0.1071 & 0.1130 \\\hline
Statistical uncertainty & $\pm 0.0019\phantom{\pm}$ & $\pm 0.0017\phantom{\pm}$ & $\pm 0.0015\phantom{\pm}$ 
                  & $\pm 0.0017\phantom{\pm}$ & $\pm 0.0023\phantom{\pm}$ \\
Systematic uncertainty  & $\pm 0.0015\phantom{\pm}$ & $\pm 0.0014\phantom{\pm}$ & $\pm 0.0015\phantom{\pm}$ 
                  & $\pm 0.0013\phantom{\pm}$ & $\pm 0.0024\phantom{\pm}$ \\ \hline
Overall experimental uncertainty 
                  & $\pm 0.0024\phantom{\pm}$ & $\pm 0.0022\phantom{\pm}$ & $\pm 0.0021\phantom{\pm}$ 
                  & $\pm 0.0021\phantom{\pm}$ & $\pm 0.0033\phantom{\pm}$ \\ \hline
Overall theoretical uncertainty 
                  & $\pm 0.0056\phantom{\pm}$ & $\pm 0.0039\phantom{\pm}$ & $\pm 0.0065\phantom{\pm}$ 
                  & $\pm 0.0062\phantom{\pm}$ & $\pm 0.0056\phantom{\pm}$ \\ \hline
$\chi^{2}$/d.o.f. & 2.2 / 9 & 10.1 / 13 & 20.2 / 11 &  9.7 / 12 & 3.8 / 8\\
\hline\hline
$\as$(200.2 \GeV)    & 0.1178 & 0.1114 & 0.1164 & 0.1088 & 0.1147 \\\hline
Statistical uncertainty & $\pm 0.0019\phantom{\pm}$ & $\pm 0.0017\phantom{\pm}$ & $\pm 0.0015\phantom{\pm}$ 
                  & $\pm 0.0017\phantom{\pm}$ & $\pm 0.0024\phantom{\pm}$ \\
Systematic uncertainty  & $\pm 0.0027\phantom{\pm}$ & $\pm 0.0028\phantom{\pm}$ & $\pm 0.0018\phantom{\pm}$ 
                  & $\pm 0.0014\phantom{\pm}$ & $\pm 0.0016\phantom{\pm}$ \\ \hline
Overall experimental uncertainty 
                  & $\pm 0.0033\phantom{\pm}$ & $\pm 0.0033\phantom{\pm}$ & $\pm 0.0023\phantom{\pm}$ 
                  & $\pm 0.0022\phantom{\pm}$ & $\pm 0.0029\phantom{\pm}$ \\ \hline
Overall theoretical uncertainty 
                  & $\pm 0.0059\phantom{\pm}$ & $\pm 0.0034\phantom{\pm}$ & $\pm 0.0062\phantom{\pm}$ 
                  & $\pm 0.0062\phantom{\pm}$ & $\pm 0.0057\phantom{\pm}$ \\ \hline
$\chi^{2}$/d.o.f. & 7.3 / 9 & 6.8 / 11 &  9.6 / 11 & 10.3 / 12 & 2.9 / 8 \\
\hline \hline
$\as$(206.2 \GeV)    & 0.1173 & 0.1119 & 0.1163 & 0.1077 & 0.1130 \\\hline
Statistical uncertainty & $\pm 0.0014\phantom{\pm}$ & $\pm 0.0013\phantom{\pm}$ & $\pm 0.0012\phantom{\pm}$ 
                  & $\pm 0.0014\phantom{\pm}$ & $\pm 0.0019\phantom{\pm}$ \\
Systematic uncertainty  & $\pm 0.0016\phantom{\pm}$ & $\pm 0.0014\phantom{\pm}$ & $\pm 0.0017\phantom{\pm}$ 
                  & $\pm 0.0013\phantom{\pm}$ & $\pm 0.0021\phantom{\pm}$ \\ \hline
Overall experimental uncertainty 
                  & $\pm 0.0021\phantom{\pm}$ & $\pm 0.0019\phantom{\pm}$ & $\pm 0.0021\phantom{\pm}$ 
                  & $\pm 0.0019\phantom{\pm}$ & $\pm 0.0028\phantom{\pm}$ \\ \hline
Overall theoretical uncertainty 
                  & $\pm 0.0057\phantom{\pm}$ & $\pm 0.0034\phantom{\pm}$ & $\pm 0.0065\phantom{\pm}$ 
                  & $\pm 0.0062\phantom{\pm}$ & $\pm 0.0053\phantom{\pm}$ \\ \hline
$\chi^{2}$/d.o.f. & 7.5 / 9  &  7.7 / 13 &  5.9 / 11 &  7.8 / 12 & 3.4 / 8\\
\hline \end{tabular}}
\end{center}
\caption[] {Values of $\as$ measured at $<\rs>$ = 194.4, 200.2 and 206.2 \GeV\ 
            from fits of the event shape variables. The fit 
            ranges, the estimated experimental and theoretical uncertainties and 
            the $\chi^{2}$/d.o.f. of the fit are also given.}
\label{tab:als}
\end{table}

\begin{table}[htbp]
\begin{center}
\begin{tabular}{|c|l@{$\pm$}r@{$\pm$}r|l@{$\pm$}r@{$\pm$}r|l@{$\pm$}r@{$\pm$}r|}
\hline
 $<\rs>$  & \multicolumn{9}{c|}{$\as$ measurement}  \\ \cline{2-10}
 (\GeV) &  \multicolumn{3}{c|}{from  $\bt$}  & \multicolumn{3}{c|}{from $\bw$}  & 
           \multicolumn{3}{c|}{from $C$}  \\ \hline
 \phantom{1}41.4 & 0.1401&0.0063&0.0119 & 0.1380&0.0067&0.0091 & 0.1371&0.0070&0.0102 \\
 \phantom{1}55.3 & 0.1321&0.0070&0.0099 & 0.1191&0.0072&0.0088 & 0.1197&0.0086&0.0118 \\
 \phantom{1}65.4 & 0.1354&0.0067&0.0106 & 0.1190&0.0062&0.0086 & 0.1258&0.0039&0.0108 \\
 \phantom{1}75.7 & 0.1296&0.0074&0.0097 & 0.1068&0.0065&0.0084 & 0.1143&0.0072&0.0094 \\
 \phantom{1}82.3 & 0.1270&0.0079&0.0095 & 0.1083&0.0067&0.0087 & 0.1153&0.0060&0.0091 \\
 \phantom{1}85.1 & 0.1259&0.0069&0.0095 & 0.1092&0.0080&0.0091 & 0.1115&0.0045&0.0089 \\
 \phantom{1}91.2 & 0.1222&0.0020&0.0080 & 0.1196&0.0022&0.0052 & 0.1170&0.0016&0.0076 \\
130.1 & 0.1178&0.0033&0.0064 & 0.1089&0.0031&0.0088 & 0.1151&0.0040&0.0066 \\
136.1 & 0.1166&0.0035&0.0064 & 0.1072&0.0041&0.0078 & 0.1089&0.0047&0.0076 \\
161.3 & 0.1123&0.0042&0.0067 & 0.1058&0.0059&0.0068 & 0.1043&0.0060&0.0057 \\
172.3 & 0.1105&0.0063&0.0061 & 0.1062&0.0050&0.0065 & 0.1121&0.0068&0.0057 \\
182.8 & 0.1145&0.0022&0.0060 & 0.1045&0.0021&0.0071 & 0.1081&0.0029&0.0054 \\
188.6 & 0.1153&0.0018&0.0067 & 0.1063&0.0017&0.0078 & 0.1118&0.0023&0.0055 \\
\hline\end{tabular}
\end{center}
\caption[]{Updated $\as$ measurements  from
           the jet broadening distributions and the $C$-parameter
           measured at $\rs <$ 189 \GeV.   The first  uncertainty is experimental 
           and the second theoretical.}
\label{tab:alscomb1}
\end{table}

\begin{table}[htbp]
\begin{center}
\begin{tabular}{|c|c|c|c|c|c|}
\hline
 $<\rs>$  & \multicolumn{5}{c|}{$\as$ measurement from $T$, $\rho$, $\bt$, $\bw$, $C$}  \\ \cline{2-6}
 (\GeV)  &   $\as$   &     stat & syst & hadr. & hi. order  \\ \hline
 \phantom{1}41.4 & 0.1418&$\pm$0.0053&$\pm$0.0030&$\pm$0.0055&$\pm$0.0085 \\
 \phantom{1}55.3 & 0.1260&$\pm$0.0047&$\pm$0.0056&$\pm$0.0066&$\pm$0.0062 \\
 \phantom{1}65.4 & 0.1331&$\pm$0.0032&$\pm$0.0042&$\pm$0.0059&$\pm$0.0064 \\
 \phantom{1}75.7 & 0.1204&$\pm$0.0024&$\pm$0.0059&$\pm$0.0060&$\pm$0.0053 \\
 \phantom{1}82.3 & 0.1184&$\pm$0.0028&$\pm$0.0053&$\pm$0.0060&$\pm$0.0051 \\
 \phantom{1}85.1 & 0.1152&$\pm$0.0037&$\pm$0.0051&$\pm$0.0060&$\pm$0.0055 \\
 \phantom{1}91.2 & 0.1210&$\pm$0.0008&$\pm$0.0017&$\pm$0.0040&$\pm$0.0052 \\
130.1 & 0.1138&$\pm$0.0033&$\pm$0.0021&$\pm$0.0031&$\pm$0.0046 \\
136.1 & 0.1121&$\pm$0.0039&$\pm$0.0019&$\pm$0.0038&$\pm$0.0045 \\
161.3 & 0.1051&$\pm$0.0048&$\pm$0.0026&$\pm$0.0026&$\pm$0.0044 \\
172.3 & 0.1099&$\pm$0.0052&$\pm$0.0026&$\pm$0.0024&$\pm$0.0048 \\
182.8 & 0.1096&$\pm$0.0022&$\pm$0.0010&$\pm$0.0023&$\pm$0.0044 \\
188.6 & 0.1122&$\pm$0.0014&$\pm$0.0012&$\pm$0.0022&$\pm$0.0045 \\
194.4 & 0.1123&$\pm$0.0018&$\pm$0.0016&$\pm$0.0020&$\pm$0.0047\\
200.2 & 0.1138&$\pm$0.0018&$\pm$0.0021&$\pm$0.0020&$\pm$0.0046\\
206.2 & 0.1132&$\pm$0.0014&$\pm$0.0016&$\pm$0.0019&$\pm$0.0047\\
\hline\end{tabular}
\end{center}
\caption[]{Combined $\as$ values   from  the five
           event shape variables with their   uncertainties.
             }
\label{tab:alscomb2}
\end{table}

\newpage

\end{document}

%% file: namelist251.tex
\typeout{   }     
\typeout{Using author list for paper 251 -- ? }
\typeout{$Modified: Jul 15 2001 by smele $}
\typeout{!!!!  This should only be used with document option a4p!!!!}
\typeout{   }
%
%
%
%
%
%

\newcount\tutecount  \tutecount=0
\def\tutenum#1{\global\advance\tutecount by 1 \xdef#1{\the\tutecount}}
\def\tute#1{$^{#1}$}
\tutenum\aachen            
\tutenum\nikhef            
\tutenum\mich              
\tutenum\lapp              
\tutenum\basel             
\tutenum\lsu               
\tutenum\beijing           
\tutenum\berlin            
\tutenum\bologna           
\tutenum\tata              
\tutenum\ne                
\tutenum\bucharest         
\tutenum\budapest          
\tutenum\mit               
\tutenum\panjab            
\tutenum\debrecen          
\tutenum\florence          
\tutenum\cern              
\tutenum\wl                
\tutenum\geneva            
\tutenum\hefei             
\tutenum\lausanne          
\tutenum\lyon              
\tutenum\madrid            
\tutenum\florida           
\tutenum\milan             
\tutenum\moscow            
\tutenum\naples            
\tutenum\cyprus            
\tutenum\nymegen           
\tutenum\caltech           
\tutenum\perugia           
\tutenum\peters            
\tutenum\cmu               
\tutenum\potenza           
\tutenum\prince            
\tutenum\riverside         
\tutenum\rome              
\tutenum\salerno           
\tutenum\ucsd              
\tutenum\sofia             
\tutenum\korea             
\tutenum\purdue            
\tutenum\psinst            
\tutenum\zeuthen           
\tutenum\eth               
\tutenum\hamburg           
\tutenum\taiwan            
\tutenum\tsinghua          

{
\parskip=0pt
\noindent
{\bf The L3 Collaboration:}
\ifx\selectfont\undefined
 \baselineskip=10.8pt
 \baselineskip\baselinestretch\baselineskip
 \normalbaselineskip\baselineskip
 \ixpt
\else
 \fontsize{9}{10.8pt}\selectfont
\fi
\medskip
\tolerance=10000
\hbadness=5000
\raggedright
\hsize=162truemm\hoffset=0mm
\def\r{\rlap,}
\noindent

P.Achard\r\tute\geneva\ 
O.Adriani\r\tute{\florence}\ 
M.Aguilar-Benitez\r\tute\madrid\ 
J.Alcaraz\r\tute{\madrid,\cern}\ 
G.Alemanni\r\tute\lausanne\
J.Allaby\r\tute\cern\
A.Aloisio\r\tute\naples\ 
M.G.Alviggi\r\tute\naples\
H.Anderhub\r\tute\eth\ 
V.P.Andreev\r\tute{\lsu,\peters}\
% ...
\newpage
\begin{list}{A}{\itemsep=0pt plus 0pt minus 0pt\parsep=0pt plus 0pt minus 0pt
                \topsep=0pt plus 0pt minus 0pt}
\item[\aachen]
 I. Physikalisches Institut, RWTH, D-52056 Aachen, FRG$^{\S}$\\
 III. Physikalisches Institut, RWTH, D-52056 Aachen, FRG$^{\S}$
\item[\nikhef] National Institute for High Energy Physics, NIKHEF, 
     and University of Amsterdam, NL-1009 DB Amsterdam, The Netherlands
\item[\mich] University of Michigan, Ann Arbor, MI 48109, USA
\item[\lapp] Laboratoire d'Annecy-le-Vieux de Physique des Particules, 
     LAPP,IN2P3-CNRS, BP 110, F-74941 Annecy-le-Vieux CEDEX, France
\item[\basel] Institute of Physics, University of Basel, CH-4056 Basel,
     Switzerland
\item[\lsu] Louisiana State University, Baton Rouge, LA 70803, USA
\item[\beijing] Institute of High Energy Physics, IHEP, 
  100039 Beijing, China$^{\triangle}$ 
\item[\berlin] Humboldt University, D-10099 Berlin, FRG$^{\S}$
\item[\bologna] University of Bologna and INFN-Sezione di Bologna, 
     I-40126 Bologna, Italy
\item[\tata] Tata Institute of Fundamental Research, Mumbai (Bombay) 400 005, India
\item[\ne] Northeastern University, Boston, MA 02115, USA
\item[\bucharest] Institute of Atomic Physics and University of Bucharest,
     R-76900 Bucharest, Romania
\item[\budapest] Central Research Institute for Physics of the 
     Hungarian Academy of Sciences, H-1525 Budapest 114, Hungary$^{\ddag}$
\item[\mit] Massachusetts Institute of Technology, Cambridge, MA 02139, USA
\item[\panjab] Panjab University, Chandigarh 160 014, India.
\item[\debrecen] KLTE-ATOMKI, H-4010 Debrecen, Hungary$^\P$
\item[\florence] INFN Sezione di Firenze and University of Florence, 
     I-50125 Florence, Italy
\item[\cern] European Laboratory for Particle Physics, CERN, 
     CH-1211 Geneva 23, Switzerland
\item[\wl] World Laboratory, FBLJA  Project, CH-1211 Geneva 23, Switzerland
\item[\geneva] University of Geneva, CH-1211 Geneva 4, Switzerland
\item[\hefei] Chinese University of Science and Technology, USTC,
      Hefei, Anhui 230 029, China$^{\triangle}$
\item[\lausanne] University of Lausanne, CH-1015 Lausanne, Switzerland
\item[\lyon] Institut de Physique Nucl\'eaire de Lyon, 
     IN2P3-CNRS,Universit\'e Claude Bernard, 
     F-69622 Villeurbanne, France
\item[\madrid] Centro de Investigaciones Energ{\'e}ticas, 
     Medioambientales y Tecnol\'ogicas, CIEMAT, E-28040 Madrid,
     Spain${\flat}$ 
\item[\florida] Florida Institute of Technology, Melbourne, FL 32901, USA
\item[\milan] INFN-Sezione di Milano, I-20133 Milan, Italy
\item[\moscow] Institute of Theoretical and Experimental Physics, ITEP, 
     Moscow, Russia
\item[\naples] INFN-Sezione di Napoli and University of Naples, 
     I-80125 Naples, Italy
\item[\cyprus] Department of Physics, University of Cyprus,
     Nicosia, Cyprus
\item[\nymegen] University of Nijmegen and NIKHEF, 
     NL-6525 ED Nijmegen, The Netherlands
\item[\caltech] California Institute of Technology, Pasadena, CA 91125, USA
\item[\perugia] INFN-Sezione di Perugia and Universit\`a Degli 
     Studi di Perugia, I-06100 Perugia, Italy   
\item[\peters] Nuclear Physics Institute, St. Petersburg, Russia
\item[\cmu] Carnegie Mellon University, Pittsburgh, PA 15213, USA
\item[\potenza] INFN-Sezione di Napoli and University of Potenza, 
     I-85100 Potenza, Italy
\item[\prince] Princeton University, Princeton, NJ 08544, USA
\item[\riverside] University of Californa, Riverside, CA 92521, USA
\item[\rome] INFN-Sezione di Roma and University of Rome, ``La Sapienza",
     I-00185 Rome, Italy
\item[\salerno] University and INFN, Salerno, I-84100 Salerno, Italy
\item[\ucsd] University of California, San Diego, CA 92093, USA
\item[\sofia] Bulgarian Academy of Sciences, Central Lab.~of 
     Mechatronics and Instrumentation, BU-1113 Sofia, Bulgaria
\item[\korea]  The Center for High Energy Physics, 
     Kyungpook National University, 702-701 Taegu, Republic of Korea
\item[\purdue] Purdue University, West Lafayette, IN 47907, USA
\item[\psinst] Paul Scherrer Institut, PSI, CH-5232 Villigen, Switzerland
\item[\zeuthen] DESY, D-15738 Zeuthen, 
     FRG
\item[\eth] Eidgen\"ossische Technische Hochschule, ETH Z\"urich,
     CH-8093 Z\"urich, Switzerland
\item[\hamburg] University of Hamburg, D-22761 Hamburg, FRG
\item[\taiwan] National Central University, Chung-Li, Taiwan, China
\item[\tsinghua] Department of Physics, National Tsing Hua University,
      Taiwan, China
\item[\S]  Supported by the German Bundesministerium 
        f\"ur Bildung, Wissenschaft, Forschung und Technologie
\item[\ddag] Supported by the Hungarian OTKA fund under contract
numbers T019181, F023259 and T024011.
\item[\P] Also supported by the Hungarian OTKA fund under contract
  number T026178.
\item[$\flat$] Supported also by the Comisi\'on Interministerial de Ciencia y 
        Tecnolog{\'\i}a.
\item[$\sharp$] Also supported by CONICET and Universidad Nacional de La Plata,
        CC 67, 1900 La Plata, Argentina.
\item[$\triangle$] Supported by the National Natural Science
  Foundation of China.
\end{list}
}
\vfill
